\documentclass[conference]{IEEEtran}
\IEEEoverridecommandlockouts

\usepackage{cite}
\usepackage{amsmath,amssymb,amsfonts}
\usepackage{algorithmic}
\usepackage{graphicx}
\usepackage{textcomp}
\usepackage{xcolor}
\usepackage{booktabs}

\usepackage{bm}

\usepackage{tabularx, booktabs, multirow}
\newcolumntype{Y}{>{\centering\arraybackslash}X}
\newcolumntype{?}{!{\vrule width 1pt}}
\usepackage{hhline}

\usepackage{listings}
\makeatletter 
\newcommand{\linebreakand}{%
  \end{@IEEEauthorhalign}
  \hfill\mbox{}\par
  \mbox{}\hfill\begin{@IEEEauthorhalign}
}
\makeatother 

\def\BibTeX{{\rm B\kern-.05em{\sc i\kern-.025em b}\kern-.08em
    T\kern-.1667em\lower.7ex\hbox{E}\kern-.125emX}}
\begin{document}


\title{First Galileo SAS Authenticated Time Solution\\
\thanks{This research was partially funded by the Research Foundation Flanders (FWO) Frank De Winne PhD Fellowship, project number 1SH9424N (Aleix Galan-Figueras).}
}


\author{
\IEEEauthorblockN{
Aleix Galan-Figueras\IEEEauthorrefmark{1}, Ignacio Fernandez-Hernandez\IEEEauthorrefmark{1}\IEEEauthorrefmark{2}, Wim De Wilde\IEEEauthorrefmark{3}, Rafael Terris-Gallego\IEEEauthorrefmark{4},\\
Gonzalo Seco-Granados\IEEEauthorrefmark{4}, Cillian O'Driscoll\IEEEauthorrefmark{5}, Sibren De Bast\IEEEauthorrefmark{3} and Sofie Pollin\IEEEauthorrefmark{1}
}
\IEEEauthorblockA{\IEEEauthorrefmark{1}KU Leuven; Leuven, Belgium}
\IEEEauthorblockA{\IEEEauthorrefmark{2}DG DEFIS, European Commission; Brussels, Belgium}
\IEEEauthorblockA{\IEEEauthorrefmark{3}Septentrio, part of Hexagon; Leuven, Belgium}
\IEEEauthorblockA{\IEEEauthorrefmark{4}Univ. Autonoma de Barcelona (UAB), CERES, IEEC; Barcelona, Spain}
\IEEEauthorblockA{\IEEEauthorrefmark{5}Independent consultant; Cork, Ireland}
}

\maketitle

\begin{abstract}
Spoofing attacks against civilian GNSS receivers have grown more common, especially near conflict zones where they now disrupt civil aviation, maritime operations, and critical infrastructure on a daily basis. Spoofing is possible because legacy civil GNSS signals are largely predictable in both their navigation data and ranging codes, allowing an attacker to forge a signal that imposes a false position and time on an unsuspecting receiver. Cryptographic authentication schemes such as Galileo's Open Service Navigation Message Authentication (OSNMA) mitigate this threat by verifying the authenticity of the navigation data. The ranging code itself, however, remains unprotected. To close this gap, Galileo is introducing a Signal Authentication Service (SAS) in the E6-C signal, which directly authenticates ranging measurements. SAS is currently transmitted by only two satellites in an elliptical orbital plane, of which at most one is visible at a time, meaning a full position solution is not yet possible; however, a georeferenced receiver can still obtain an authenticated time solution. This paper presents, to the authors' knowledge, for the first time, a timing solution computed from an authenticated civil GNSS signal. We develop a snapshot software receiver implementing a simplified version of the Galileo SAS protocol to compute the receiver clock bias from an authenticated pseudorange, using radio-frequency data recorded with an engineering prototype software-defined radio receiver from Septentrio. We evaluate the resulting timing solution using recordings from both SAS-capable satellites collected at different locations, demonstrating the feasibility of authenticated timing ahead of full SAS operational deployment.
\end{abstract}



\section{Introduction}

In recent years, the problem of Global Navigation Satellite Systems (GNSS) interference affecting civilian operations has increased. The once sporadic occurrence can now be experienced daily by planes, ships, and critical civilian infrastructure near conflict areas \cite{menzione_interference_2025, gpsjam, gpswise}. This interference includes spoofing attacks that forge GNSS signals and transmit them to mislead unaware receivers into reporting incorrect position and time.

Several solutions have been proposed during the past decade to tackle spoofing. These range from simple receiver-based consistency checks, such as monitoring sudden changes in the received carrier-to-noise ratio ($C/N_0$) and automatic gain control, to more complex signal-processing techniques. Multi-antenna receivers can estimate the direction of arrival of incoming signals and employ adaptive beamforming or null-steering algorithms to suppress signals arriving from suspicious directions \cite{spoofing_and_detection}. Although these methods significantly increase the difficulty of a successful attack, they often require additional hardware, external sensors, or sophisticated algorithms, and they cannot always distinguish a subtle spoofer from authentic signals with complete certainty.

A fundamentally different approach is to protect the signal at its source through cryptographic authentication. Rather than attempting to detect whether a received signal appears suspicious, the receiver can mathematically verify that it is genuine. While this method is another strong layer of security, it comes with its own drawbacks, the main one being a latency in the authentication. The European Galileo system is currently the first GNSS to operationally deploy such a capability through its Open Service Navigation Message Authentication (OSNMA) service \cite{OSNMA_ICD}.

The OSNMA service was declared operational in July 2025 and is transmitted in the E1-B signal to authenticate the I/NAV navigation messages broadcast on the E1-B and E5b signals. It does so by transmitting cryptographic authentication tags as part of the navigation message. The symmetric cryptographic key used to generate the tags is intentionally disclosed only after the tags have been broadcast. Consequently, a spoofer cannot forge valid authentication tags in advance and is therefore unable to modify the navigation data. This delayed key disclosure mechanism is based on an adaptation of the Timed Efficient Stream Loss-Tolerant Authentication (TESLA) protocol for Galileo \cite{perrig2003tesla}, in which each disclosed key is authenticated through a one-way hash chain anchored at a trusted root key. The root key itself is authenticated using a long-term public key \cite{fernandez2016navigation}.

However, the actual ranges of the signal used to compute the GNSS solution are still unprotected. Receivers can use the unpredictability of OSNMA bits, but this has limitations, as only the first samples of each unpredictable symbol are unpredictable \cite{o2022mapping} \cite{seco2021detection}. Galileo is deploying a range authentication protocol named Signal Authentication Service (SAS) using the E6-C signal. In this initial phase, Galileo SAS is transmitted only by the two satellites in an elliptical orbit (space vehicle identifiers E14 and E18), which are 180 degrees apart in their common orbital plane, so only one of them is visible from the Earth's surface at a given epoch. With one satellite in view transmitting SAS, computing a PVT solution is not feasible, but with a georeferenced antenna, it is possible to compute an authenticated time solution.

Proposals for authenticated timing using open signals have emerged in recent years \cite{wesson2013probabilistic}. A theoretical framework for exploiting multiple Galileo SAS observations to bound and monitor the timing solution was introduced in \cite{ardizzon2022authenticated}. More recently, \cite{anderson2025world} authenticated the pseudorange from the Pulsar-O low Earth orbit satellite. To date, however, no authenticated timing solution has been demonstrated using actual Galileo
SAS measurements.

Although Galileo SAS (formerly known as ACAS, Assisted Commercial Authentication Service) is not yet operational, there has been growing interest within the GNSS community in GNSS signal authentication \cite{scott2003anti}, and SAS in particular \cite{schreiber2025range} \cite{dorinsvalidation}. An open-source Python implementation of a SAS snapshot-processing library has been made available on GitHub \cite{galileosaslib}. The library is actively being developed and is expanding its features, but it currently does not support Position-Velocity-Time (PVT) solutions.

In this paper, we present and analyze, to our knowledge, the first timing solution using an authenticated civil GNSS signal. We use a self-developed snapshot software receiver that implements a simple version of the Galileo SAS protocol and uses authenticated ranges to compute the receiver clock bias. To record the radio-frequency signal, we use an engineering prototype receiver from Septentrio that incorporates a software-defined radio (SDR). Finally, we analyze the timing solution with recordings from both satellites in different locations.


\section{Galileo SAS}

\subsection{Protocol overview}

The Galileo SAS protocol \cite{fernandez-hernandez_galileo_2024} authenticates the ranges by transmitting unpredictable spreading code sequences in the E6-C signal, whose content is disclosed a posteriori. A receiver records the RF signal during the unpredictable sequence period and verifies, by means of correlation, that the received signal was authentic.

To obtain an unpredictable signal, the Galileo E6-C signal spreading code is encrypted with a secret key. Currently, the E6-C signal is used as a complementary pilot signal for the Galileo High Accuracy Service (HAS) provided in E6-B. Before SAS is declared operational, the E6-C will be encrypted permanently, and it will no longer be possible to use the signal as a pilot until a revised signal plan is provided in Galileo 2nd Generation.

Sections of the encrypted E6-C are selected for disclosure to users. These sections, called Encrypted Code Sequences (ECS), are up to 16 ms long and transmitted every 200 ms, synchronized with Galileo System Time (GST). Galileo SAS discloses different ECS slots within the 200 ms period for different use cases; for our experiment, we will use the first slot (i.e., the first 16 ms).

Naturally, disclosing the ECS in advance would allow spoofers to re-create the signal, defeating the purpose of range authentication. Therefore, the ECS are re-encrypted using the nearest future TESLA key from OSNMA, thereby creating Re-Encrypted Code Sequences (RECS). The RECS can be safely disclosed in advance, as they will only be decrypted after they have been transmitted in the E6-C signal. The RECS are stored on a public server and can be downloaded beforehand.

\begin{figure}
\centerline{\includegraphics[width=0.5\textwidth]{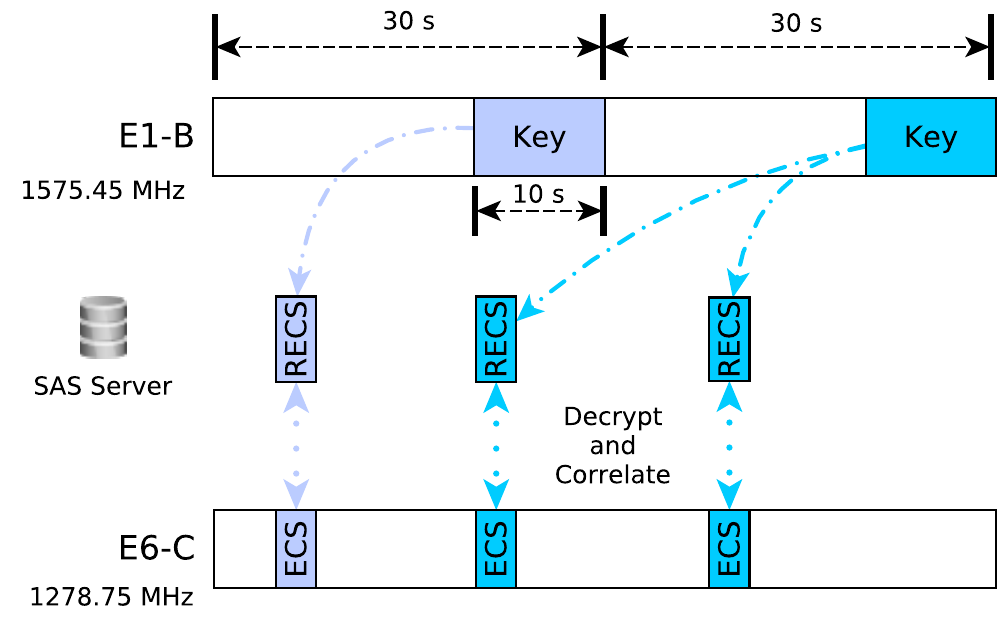}}
\caption{Simplified Galileo SAS diagram. A receiver implementing SAS must download the RECS in advance. When the TESLA key is disclosed in E1-B and authenticated, it can decrypt the RECS and correlate against the RF snapshot containing the ECS.}
\label{fig:sas_operation}
\end{figure}

A receiver implementing Galileo SAS must first download and store RECS from the SAS server for the desired autonomy period. Then, it must record an RF snapshot when the corresponding ECS is expected to be received. After the TESLA key is disclosed in E1-B and authenticated, it can decrypt the RECS and correlate with the recorded snapshot to detect the presence of the ECS. This process is depicted in Fig. \ref{fig:sas_operation}, and further explained in \cite{semiassisted}.

To authenticate a PVT solution, the receiver may compute a solution directly using the results of the E6-C correlation. Alternatively, it may translate the code delay from E6-C to an open signal that is currently in tracking (for example, E1-B) and evaluate whether they are close enough to consider those ranges authentic.

Finally, it is important to note that Galileo SAS has inherent authentication latency. The last ECS for which a TESLA key is applicable is transmitted just before the key. In the minimum-latency scenario, a receiver wants to authenticate the ECS just before the TESLA key, which incurs a latency of 10 seconds (Fig. \ref{fig:sas_operation}). In any case, the Time Between Authentications (TBA) will be 30 seconds: the time until a new TESLA key is transmitted. 

\subsection{Galileo SAS in a snapshot receiver}

A snapshot receiver does not continuously track the GNSS signal to obtain measurements and navigation data. Instead, it takes short RF snapshots at a low duty cycle and calculates ranges based on their contents. Because the snapshots are not long enough to decode any navigation data, the receiver must obtain the ephemeris externally, for example, via downloaded RINEX files. For Galileo SAS, in addition to the ephemeris, the receiver needs authenticated TESLA keys to decode the RECS. For our proof of concept, we use an internet side channel to connect to the public OSNMAlib API and retrieve the authenticated keys \cite{galan2025improvingosnmalib}.

In our snapshot receiver, we record E6 (at 1278.75 MHz) to have access to both E6-B and E6-C. The open E6-B signal is used to obtain a fine Doppler estimate via circular correlation in a typical GNSS acquisition mode. The Doppler estimation is then used to bootstrap and reduce the computational complexity when correlating the ECS in E6-C \cite{terris2025efficient}.

It is important to keep in mind that the correlator's response to a Doppler mismatch follows a sinc shape, whose first null is at the inverse of the coherent integration time. Therefore, the selected bin spacing determines the worst-case loss. For correlating a complete ECS of 16 ms, a bin spacing of 32 Hz provides a maximum loss of approximately 1 dB \cite{borre2022gnss}.


\section{RF snapshots recording}

\subsection{Septentrio prototype SDR}

To record the RF snapshots needed for our receiver to work, we use an internal prototype firmware from Septentrio that introduces an SDR functionality to their Mosaic G5 receiver \cite{sept_g5}. The prototype SDR functionality allows us to take 50-ms snapshots every 200 ms at a 10 MHz sampling frequency with 2-bit quantization. The snapshot period is sufficient to obtain a snapshot for every RECS, and the sampling frequency is roughly sufficient to sample the main lobe of the E6-B/C signal, which is modulated in BPSK(5). Finally, since GNSS signals are below the noise floor, two bits per sample do not introduce much quantization loss and help keep the snapshot size small (four bits per IQ complex sample).

The snapshots are taken in synchronization with Galileo System Time (GST), as computed by Septentrio's receiver standard firmware based on E1 and E5 measurements, which we consider valid a priori and authenticate later with the SAS pseudorange. However, the SAS E6-C pseudorange is subject to the internal hardware delay of the channels used for the SDR feature, which we calibrate separately.

From the point of view of our snapshot receiver, we are mostly transparent to Septentrio's receiver PVT calculations. In other words, the SAS post-processing does not use the receiver’s conventional PVT solution beyond the coarse RF snapshot scheduling and timestamping. A similar setup without the prototype firmware could be replicated by using Septentrio's receiver PPS as a hardware trigger for a commercial SDR such as the BladeRF 2.0 \cite{rafa_sync_blade}.

\subsection{Snapshot recording time}

A central decision for a successful snapshot SAS receiver is when to take the RF snapshots. While the ECS transmission time is known (every 200 ms, aligned with GST), we are interested in the reception time.

\begin{figure}
\centerline{\includegraphics[width=0.3\textwidth]{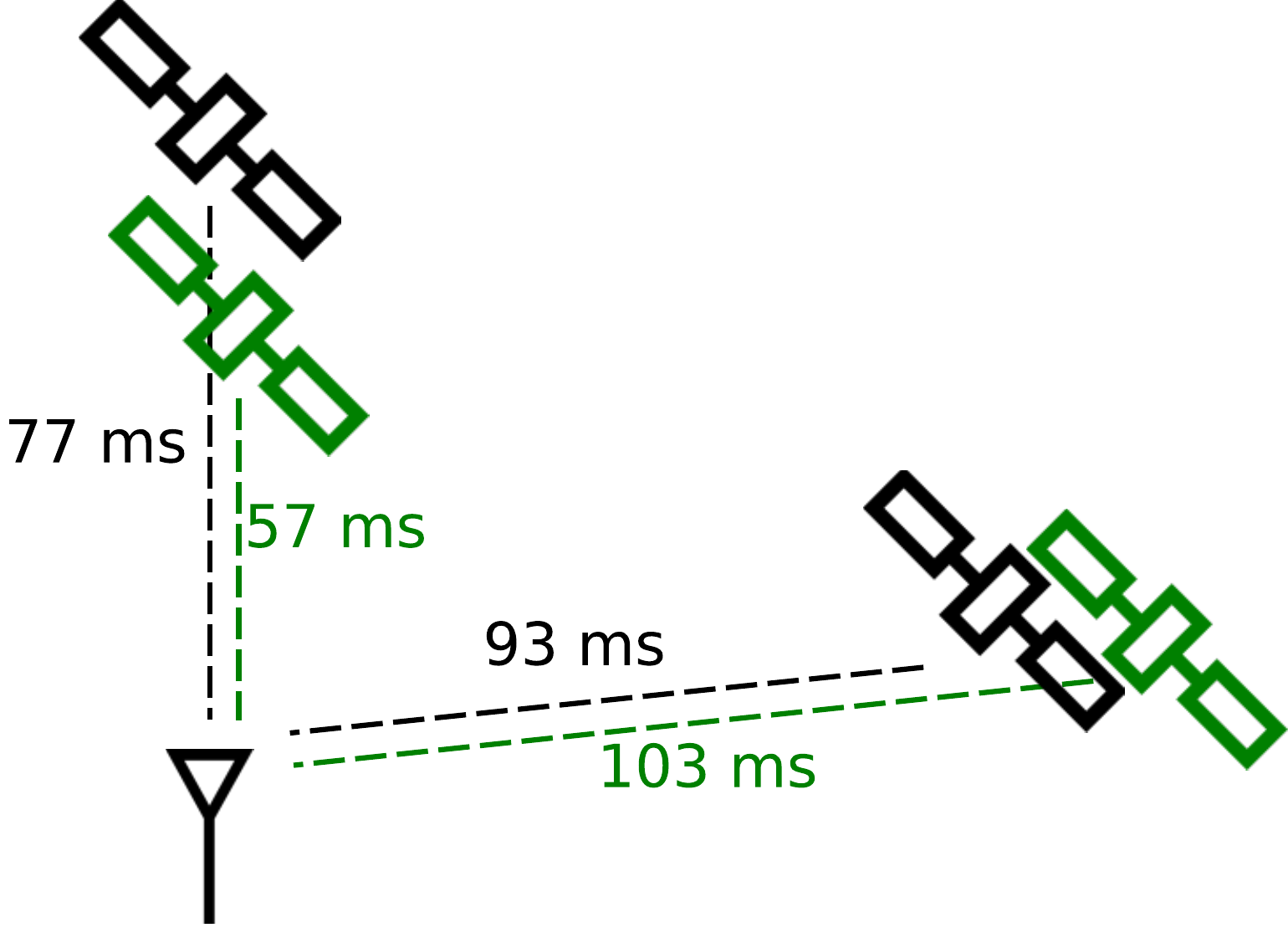}}
\caption{Signal propagation time for nominal orbit Galileo satellites (in black) and elliptic orbit satellites (in green) at the zenith and at 10 degrees of elevation.}
\label{fig:signal_propagation}
\end{figure}

For Galileo satellites in nominal orbits, the signal takes approximately 77 ms to reach the Earth's surface from a satellite at the zenith (90 degrees of elevation) and 93 ms from a satellite at 10 degrees of elevation (Fig. \ref{fig:signal_propagation}). There is a maximum 15 ms delay difference among any two satellites in the sky, which should be added to the 16 ms ECS length. If all clocks were perfect, recording for 31 ms should suffice to obtain all ECS from all satellites in view. However, the satellite clock error in Galileo satellites can reach up to ±4 ms, and receiver biases should also be accounted for, which are implementation-dependent. In the end, a 50-ms snapshot is a sound value for our setup.

Nonetheless, SAS is currently transmitted only by the satellites E14 and E18, which are in an elliptical orbit with an eccentricity of 0.162. These satellites can be as close as 57 ms (perigee at the zenith) and as far as 103 ms (apogee at 10 degrees). Since only one of them is in view at a time, this is not a problem for our 50-ms snapshots. The satellite trajectory can be roughly estimated from the ephemeris, and the snapshot can be configured to start at a sensible time that covers the entire satellite pass.

\section{Timing receivers}

\subsection{Model}

Conventional GNSS receivers solve for four unknowns simultaneously: the three components of receiver position $(x, y, z)$ and the receiver clock bias $(b)$ relative to a GNSS system time. Because of the four unknowns, at least four satellites must be in view at once. In a dedicated GNSS timing receiver, the position is already known (the antenna is fixed and precisely surveyed), so the navigation problem collapses to a single unknown: the receiver clock bias.

Let $\rho$ be the authenticated pseudorange, where $r$ is the geometrical range between the receiver position $\bm{x}$ and the satellite position $\bm{x_s}$, $b$ is the receiver clock bias (in meters), $dt$ is the satellite clock bias, $I$ is the ionospheric delay, $T$ is the tropospheric delay, and $\epsilon_\rho \sim \mathcal{N}(0,\sigma_\epsilon^2)$ models the aggregate residual error, including measurement noise, including multipath, jitter, and other effects (for convenience, its fast-varying component is approximated as zero-mean Gaussian, although slowly varying orbit, atmospheric, and multipath errors may introduce nonzero biases):

\begin{equation}
\label{eq:geometric_range}
r = \sqrt{(x_s - x)^2 + (y_s - y)^2 + (z_s - z)^2} ,
\end{equation}

\begin{equation}
\label{eq:pseudorange}
\rho = r + b - dt\cdot c + I + T + \epsilon_\rho .
\end{equation}

In this model (\ref{eq:pseudorange}), only $b$ is unknown, and therefore it can be determined from a single equation:

\begin{equation}
\label{eq:clock_bias}
\hat{b} = \rho - \hat{r} + \hat{dt}\cdot c - \hat{I} - \hat{T} .
\end{equation}

If all other parameters are authenticated, by determining $\hat{b}$, the receiver can provide a trusted time reference.

\subsection{Parameter estimation}

The pseudorange $\rho$ is estimated from the reception time $t_{rx,ECS}$ (in receiver time), and the ECS transmission time $t_{tx,ECS}^{GST}$ (in GST timescale), which is known from the downloaded RECS file:

\begin{equation}
\label{eq:measured_range}
\rho = (t_{rx,ECS} - t_{tx,ECS}^{GST}) \cdot c .
\end{equation}

The Galileo SAS signal does not repeat, allowing the measured correlation delay to be mapped uniquely to the signal transmission time, without the code phase ambiguity. Since this transmission time is derived from the encrypted E6-C spreading sequence, it constitutes an authenticated timing reference. For this work, we assume it cannot be forged without knowledge of the secret spreading code. In other words, the attacker cannot synthesize or predict the ECS before its transmission. Advanced meaconing attacks on authenticated signals that forge receiver time are outside the scope of this work. These attacks can, however, be detected or mitigated by examining received signal power and vestigial signals \cite{winkel2024combining}.

The satellite position $\bm{x_s}$ at the ECS transmission time is calculated using the ephemeris from a RINEX file obtained via a secure side channel. In a normal GNSS receiver, the ephemeris would be obtained from E1-B and authenticated with OSNMA. The receiver position $\bm{x}$ is assumed to be trusted, as it is static and georeferenced. Therefore, $r$ can be trusted.

The satellite clock bias $\hat{dt}$ is estimated using the clock drift parameters in the ephemeris (from the RINEX file) and the E6-C Broadcast Group Delay (BGD) provided by SAS (since the clock model is referred to the E1-E5b iono-free combination). In a normal receiver, the clock parameters are obtained from the I/NAV message and authenticated with OSNMA. The $dt$ term also includes the relativistic error, which is particularly significant for the elliptic satellites and is calculated internally from the authenticated satellite position.

\begin{equation}
\label{eq:sv_clock_bias}
\hat{dt} = dt_{INAV}(t_{tx,ECS}^{GST}) - BGD(t_{tx,ECS}^{GST}) .
\end{equation}

Here, \(dt_{\mathrm{I/NAV}}\) denotes the complete broadcast satellite clock correction, including both the polynomial clock model and the relativistic correction, evaluated at the estimated transmission time \(t_{tx,ECS}^{GST}\).

The ionosphere delay $\hat{I}$ is estimated using the ionosphere correction estimate obtained from the I/NAV message and the NeQuickG model. These parameters are also authenticated with OSNMA in a normal GNSS receiver. In our case, we obtain them from the RINEX file and use the NeQuickJRC library \cite{GSC_NeQuickG} to calculate the E6-C delay.

The tropospheric delay $\hat{T}$ is estimated internally from the authenticated satellite position and is therefore also trusted. In our receiver, we implement the Saastamoinen model \cite{saastamoinen1972atmospheric} with the Niell model mapping function \cite{niell1996global}.

Therefore, since all parameters used to estimate $b$ are authenticated, $\hat{b}$ can be authenticated as well. Note, however, that its accuracy remains affected by residual errors (satellite orbit and clock, atmospheric, multipath, and receiver errors). The absolute time reference (in GST) can therefore be determined as:

\begin{equation}
\label{eq:absolute_time}
t^{GST}(t_{rx,ECS}) = t_{rx,ECS} - \hat{b} / c .
\end{equation}

To estimate the time between ECS authenticated epochs, the receiver can estimate its clock drift $b'$ at $t_{rx,ECS}$ using two consecutive ECS and propagate its estimation:

\begin{equation}
\label{eq:propagated_time}
t^{GST}(t) = t^{GST}(t_{rx,ECS}) + (1 - \hat{b'} / c) \cdot (t - t_{rx,ECS}) .
\end{equation}


\section{Results}

\subsection{Receiver calibration}

To evaluate the authenticated time solution, we recorded five static scenarios: four for calibration and one for evaluation. All scenarios are recorded using a georeferenced antenna in an open-sky environment near Leuven, Belgium. The four calibration scenarios are summarized in Table \ref{tab:recorded_scenarios}. We recorded two passes of the E14 and two of the E18 at different times of the day. We use the closest ECS to the key that authenticates it. This gives a latency of 10 s, the time it takes for the TESLA key to be transmitted. We calculate one ECS for every new disclosed key, so the TBA is 30 seconds.

\begin{table}[h]
    \centering
    \caption{Calibration scenarios}
    \label{tab:recorded_scenarios}
    \begin{tabular}{@{}lccccc@{}}
        \toprule
        Scenario & Day & Start (GNSS) & Duration & \# Points & Satellite \\
        \midrule
        S1 & 2026-06-24 & 07:35:20 & 150 min & 300  & E14 \\
        S2 & 2026-07-15 & 08:27:00 & 120 min & 240  & E14 \\
        S3 & 2026-07-18 & 06:47:00 & 170 min & 340  & E18 \\
        S4 & 2026-08-24 & 21:00:00 & 300 min & 600  & E18 \\
        \bottomrule
    \end{tabular}
\end{table}

\begin{figure}[h]
\centerline{\includegraphics[width=0.5\textwidth]{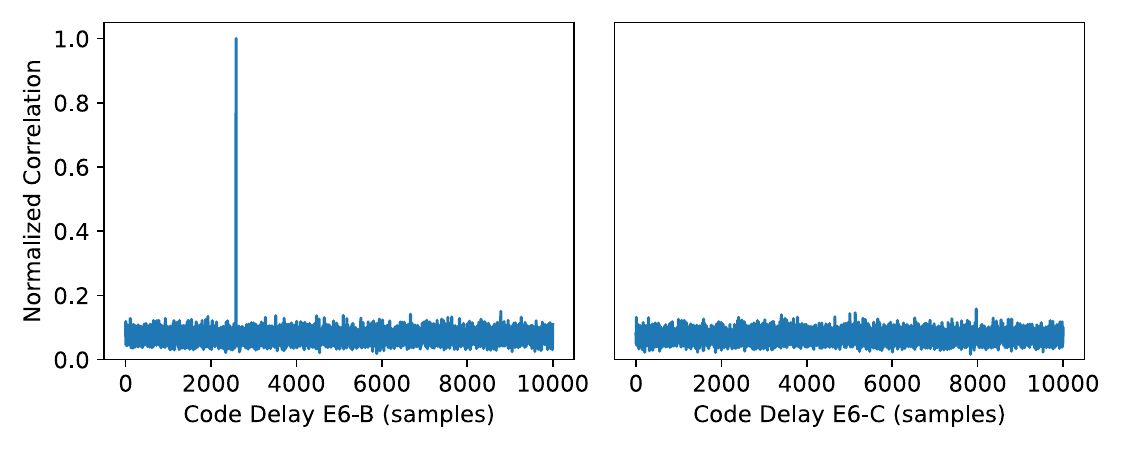}}
\caption{Result of the correlation with the public ranging codes in the E6-B and E6-C signals for the satellite E14. The signal E6-C is encrypted.}
\label{fig:acq_encrypted}
\end{figure}

\begin{figure}[h]
\centerline{\includegraphics[width=0.5\textwidth]{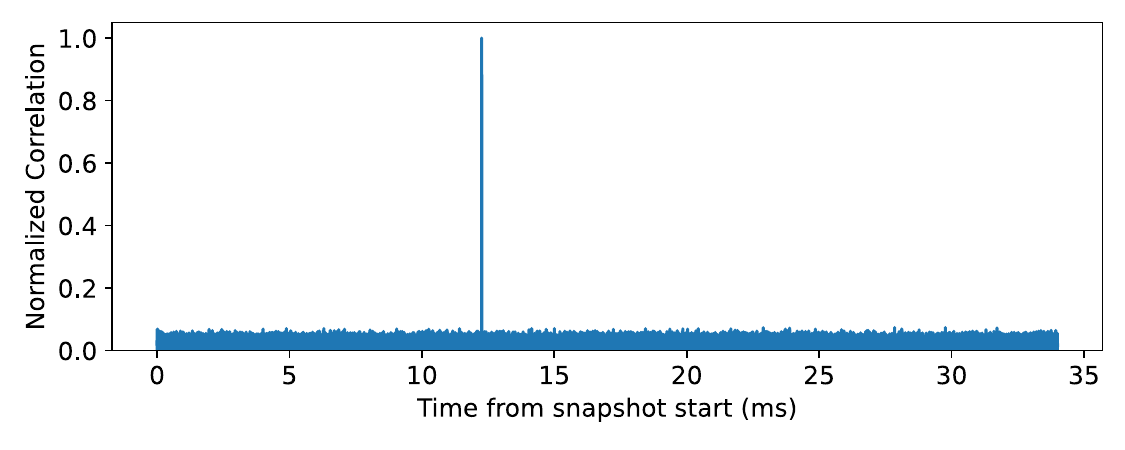}}
\caption{Result of correlating the decrypted RECS with the recorded RF snapshot containing an ECS for the satellite E14.}
\label{fig:recs_correlation}
\end{figure}

Figure \ref{fig:acq_encrypted} shows, for the satellite E14, how the E6-C signal is encrypted by correlating with the public spreading code in E6-B and E6-C, with only the correlation peak visible in E6-B. Then, Fig. \ref{fig:recs_correlation} shows the resultant peak after correlating with the decrypted ECS. We use a quadratic interpolator to obtain an accurate estimate of the peak delay, since it is usually located between two samples \cite{borre2022gnss}. The peak delay needs to be as precise as possible because it is used to calculate the pseudorange.

The calculated receiver clock bias results for the calibration scenarios are shown in Table \ref{tab:sum_results}.

\begin{table}[h]
    \centering
    \caption{Receiver clock bias results for all scenarios}
    \label{tab:sum_results}
    \begin{tabular}{@{}lccccc@{}}
        \toprule
        Scenario & Time (GNSS) & Mean (\textmu s) & SD (ns) & Max Elev (deg) \\
        \midrule
        S1 & 07:35:20 & 5.382 & 25.0 & 81.1 \\
        S2 & 08:27:00 & 5.370 & 21.2 & 62.5 \\
        S3 & 06:47:00 & 5.362 & 18.9 & 66.7 \\
        S4 & 21:00:00 & 5.359 & 16.2 & 86.0 \\
        \bottomrule
    \end{tabular}
\end{table}

The measured receiver clock bias mainly comprises three components: the receiver clock offset itself; a fixed instrumental delay arising from the uncalibrated SDR chain in the Septentrio receiver; and other time-varying errors (residual tropo/iono delay, ephemeris error, multipath, and receiver noise). The fixed instrumental delay is common to all recordings and represents a constant offset, whereas the other errors vary between scenarios.

While the mean receiver block bias across all scenarios should be similar (driven by the instrumental delay), it differs slightly. The mismatch is likely due to residual ionospheric error and the reduced number of points in the experiments. For example, scenario S1 has the satellite located in the south during a period of high ionospheric activity, which increases measurement variability as we are not using a dual-frequency receiver to remove the ionosphere component.

As discussed previously, in a fully calibrated timing receiver, the instrumental delay would be removed, leaving the receiver clock bias together with the satellite-specific errors: ephemeris orbit error, multipath, and residual atmospheric delays; plus the white measurement noise. When several satellites are combined, these per-satellite errors are largely independent and are reduced, e.g., by least-squares estimation, while the receiver clock bias, being common to all satellites, remains. For the authenticated timing signal, only a single satellite is available, so the clock bias cannot be separated from the satellite-specific errors.

In our calibration, the instrumental delay is estimated as the combined mean of all the scenario data points. Given the limited amount of measurements, the estimated instrumental bias may still contain satellite-specific errors and measurement white noise. For an operational timing receiver using Galileo SAS, the receiver hardware bias should be pre-calibrated. The estimated instrumental delay is 5.366 \textmu s, which is consistent with initial calibration results from Septentrio for the receiver SDR channels.

\subsection{Time solution evaluation}

To evaluate the time solution, we subtract the estimated instrumental delay (5.366 \textmu s) from the measurements and analyze the resulting root mean squared error (RMS), assuming it is due to the slowly varying geometry, residual ionospheric estimation errors, and ephemeris systematics in addition to the white measurement noise.

\begin{figure}[h]
\centerline{\includegraphics[width=0.5\textwidth]{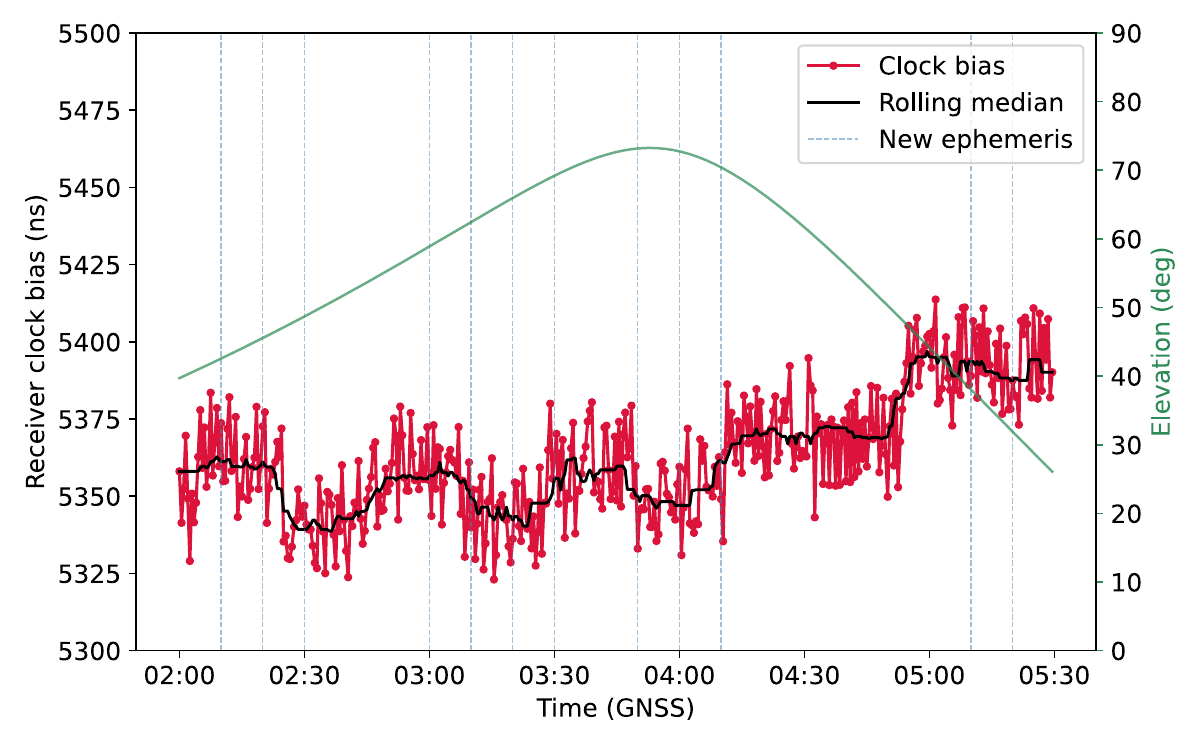}}
\caption{Receiver clock bias measured with the satellite E18 in an open sky environment on September 9th, 2026. The increasing clock bias values are due to residual ionosphere.}
\label{fig:results_bias}
\end{figure}

The evaluation scenario is a 210-minute pass of the E18 satellite on September 9th, 2026, starting at 02:00:00 GNSS time. The snapshot receiver clock bias results for this scenario are shown in Figure \ref{fig:results_bias}. The figure shows the receiver's instantaneous clock bias in red, computed from the ECS correlation pseudorange, and the trend in black, using a rolling median. We also overlap the elevation in green and the new ephemeris transmission as vertical blue lines. The increasing trend in the values is likely due to residual ionosphere, which increases with daylight (as local time is 2 hours ahead of GNSS time).

\begin{figure}[h]
\centerline{\includegraphics[width=0.5\textwidth]{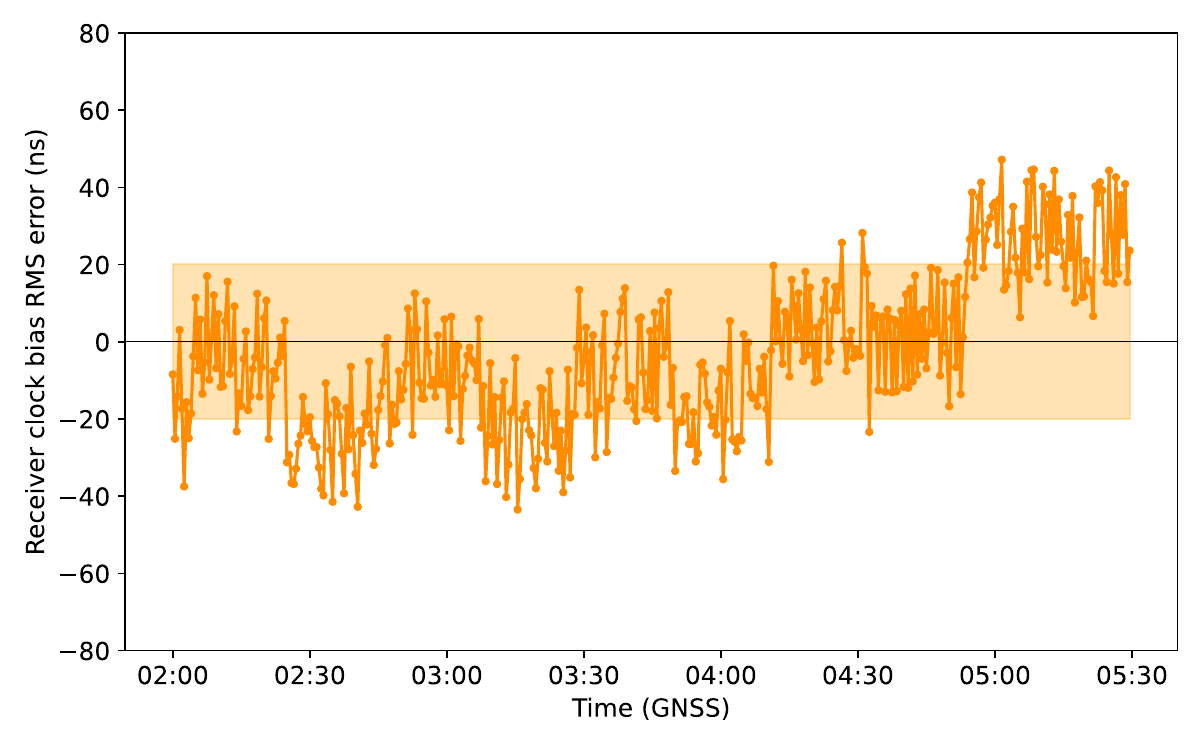}}
\caption{Receiver clock bias RMS error measured with the satellite E18 in an open sky environment on September 9th, 2026. The orange band marks ±RMS (20.0 ns).}
\label{fig:results_scatter}
\end{figure}

The RMS error after subtracting the calibrated instrumental delay is presented in Fig. \ref{fig:results_scatter}. The orange bands in the figure indicate ±RMS error of the values, which is 20.0 ns (5.99 m). These RMS results are in line with a snapshot receiver using a single satellite in single frequency.

When the receiver is properly calibrated, the RMS error can be modeled as only the residual atmospheric error, ephemeris error, and white noise. We assume that, under our attack model, an attacker cannot successfully alter the SAS signal without being unnoticed, so the measured time from the SAS correlation peak is considered authentic. Protection against sophisticated threats beyond the scope of our work can be obtained with more SAS satellites, among other measures.


\section{Conclusion}

In this paper, we presented the first authenticated time results using Galileo SAS, which is currently transmitted only by the Galileo E14 and E18 satellites in elliptical orbits.

As only one SAS measurement is needed to authenticate time in a georeferenced timing receiver, we developed a snapshot timing receiver that can process RECS from the SAS server, decrypt them using the TESLA key, and correlate them with recorded RF data to detect the presence of the ECS. The receiver uses the E6-B signal to aid Doppler estimation, and the ephemerides are provided via RINEX files. However, in the model discussion, we have shown how a receiver that obtains the ephemeris from E1-B can authenticate all necessary parameters using OSNMA.

To record RF snapshots, we have used an experimental internal firmware from a Septentrio mosaic-G5 receiver. With this feature, we can record RF samples synchronized with GST and later authenticate them using the SAS pseudorange. We account for signal propagation times when starting the RF snapshot recording to receive the ECS from all satellites in view. For the Galileo elliptical satellites, signal propagation time can vary from 57 to 103 ms.

To test the authenticated timing solution, we recorded five passes of the satellites E14 and E18 at different times. Four of the recordings are used to calibrate the receiver's hardware bias for E6-C, yielding 5.366 \textmu s. Using the hardware bias as reference, we obtain an RMS error of 20.0 ns (5.99 meters) in the last recording, which seems commensurate with typical GNSS residual errors. This proves that, after calibrating the biases, the SAS-authenticated timing results align with expectations. To the best of the authors' knowledge, this is the first experimental computation of a GNSS  timing solution from in-orbit cryptographically authenticated civil GNSS satellite ranging signals.

The next steps in the research include, among others, recording additional scenarios and data points and evaluating a full SAS PVT solution, which, even for a timing receiver, further constrains possible attacks. Also, it would be interesting to calculate the best possible accuracy by using very precise SP3 orbit files and real-time maps for the atmospheric estimates. Finally, we are integrating this snapshot-timing receiver into the open-source GalileoSASlib library.


\bibliographystyle{IEEEtran}
\bibliography{biblio.bib}

\end{document}